\documentclass[aps,prl,floatfix,twocolumn,superscriptaddress]{revtex4-1}
\usepackage{amsmath,amssymb}
\usepackage{graphicx}
\usepackage{epsfig}
\usepackage{psfrag}
\usepackage[usenames]{color}
\usepackage{xcolor}
\usepackage{hyperref}
\usepackage{soul,color}
\hypersetup{colorlinks=true,citecolor={blue},linkcolor={blue},urlcolor={blue}}

\begin{document}

\hsize\textwidth\columnwidth\hsize\csname@twocolumnfalse\endcsname

\title{Bogolon--mediated electron capture by impurities \\in hybrid Bose--Fermi systems}

\author{M.~V.~Boev}
\affiliation{A.V. Rzhanov Institute of Semiconductor Physics, Siberian Branch of Russian Academy of Sciences, Novosibirsk 630090, Russia}

\author{V.~M.~Kovalev}
\affiliation{A.V. Rzhanov Institute of Semiconductor Physics, Siberian Branch of Russian Academy of Sciences, Novosibirsk 630090, Russia}
\affiliation{Department of Applied and Theoretical Physics, Novosibirsk State Technical University, Novosibirsk 630073, Russia}

\author{I.~G.~Savenko}
\affiliation{Center for Theoretical Physics of Complex Systems, Institute for Basic Science, Daejeon, Republic of Korea}

	

\date{\today}

\begin{abstract}
We investigate the processes of electron capture by a Coulomb impurity center residing in a hybrid system consisting of spatially separated two-dimensional layers of electron and Bose-condensed dipolar exciton gases coupled via the Coulomb forces. We calculate the probability of the electron capture accompanied by the emission of a single Bogoliubov excitation (bogolon), similar to regular phonon--mediated scattering in solids. Further, we study the electron capture mediated by the emission of a pair of bogolons in a single capture event and show that these processes not only should be treated in the same order of the perturbation theory, but also they give more important contribution than single bogolon--mediated capture, in contrast with regular phonon scattering.
\end{abstract}	


\maketitle


\section{Introduction}
The presence of impurities in semiconductor nanostructures strongly modifies their physical properties~\cite{JenaPRL981368052007,GibbonsPRL1022555022009}. At low temperature, the electron-impurity scattering is predominant and it determines the electric properties of the heterostructure, in particular, its conductivity~\cite{SimonBook}. Depending on the sign of the electron-impurity interaction, electrons can be either scattered by the impurities or captured by them~\cite{EshchenkoPRL892266012002, ShiPRL1092455012012, PalmaPRB51141471995, BourgoinPRB45113241992}. In terms of the classical Drude theory, the former processes modify the effective scattering time of the electrons, whereas the latter processes literally result in the decrease of the number of free carriers of charge. As a result, non-radiative capture of electrons by charged attractive centers plays crucial role in the transport of photoexcited carriers~\cite{Abakumov}, drastically modifying the conductivity via electron lifetime. In particular, this lifetime is a crucial parameter for impurity photodetectors~\cite{Aleshkin2015, Kozlov2015}, which are commonly used in far-infrared range to monitor the emission from modern resonant tunneling diodes and quantum cascade lasers.

In the majority of cases, an electron capture is accompanied by the emission of crystal lattice excitation quanta referred to as acoustic and optical phonons~\cite{GummelPR9714691955, LaxPR11915021960, AbakYass}. Meanwhile the electron loses its energy and becomes localized. Phonon--mediated electron scattering has been so far considered to be the dominant capture mechanism.
However, lattice vibrations are not the only phonons available, especially at low temperatures. For instance, in view of recent discovery of exciton superfluidity and Bose-Einstein condensation (BEC)~\cite{RefButov}, one can consider the excitations of the BEC as an alternative type of phonons, commonly referred to as Bogoliubov quanta or bogolons and having linear dispersion law at small momenta.
Such exciton BEC can be realized experimentally by external laser beams which produce photo-excited electrons and holes, relaxing their energy to form bound electron-hole pairs. We will show that in the presence of exciton gas, the interaction of the carriers of charge with impurities can be strongly modified, if the exciton gas is in the BEC phase.

In order to better understand fundamental properties of this phenomenon, it is important to separate the BEC from the conduction electrons and study the influence of different interactions separately. One of the recent active areas of research is hybrid Bose-Fermi systems which consist of two-dimensional (2D) spatially separated electron and exciton gases, interacting with each other via the Coulomb forces~\cite{RefOurPRB, RefSR, Cotlet, Laus, Matus}. These systems can be a testbed for various physical phenomena, some of which occur when the exciton or exciton-polariton gas is in the BEC regime~\cite{KovalevJETP1, KovalevJETP2, KovalevJETP3} which has been reported in various solid state systems~\cite{RefButov, Kasprzak, OurNature}. In particular, the possibility of inelastic processes of electron capture has been so far disregarded, to the best of our knowledge.

In this article we will demonstrate that in the presence of exciton BEC, an additional mechanisms of electron capture to attractive centers appears. This mechanism is the consequence of interlayer electron-exciton interaction. Being in the BEC regime, exciton gas can be described in terms of bogolons (with a soundlike dispersion in the long--wavelength limit). Naively, one can expect that the processes of electron capture due to interaction with the BEC of excitons are similar to the case of lattice phonon emission, in particular, due to the similarity of the dispersion laws. Indeed, it is partly true. However, we will show that in the presence of the BEC, an additional channel of non-radiative relaxation of electrons opens. It can be referred to as electron capture accompanied by the radiation of a pair of bogolons. Counterintuitively, such electron capture events should be treated within the same order of perturbation theory as the single bogolon emission, moreover, as it will be demonstrated, they give more important contribution.


\section{System schematic}
\begin{figure}[!t]
	\includegraphics[width=0.7\linewidth]{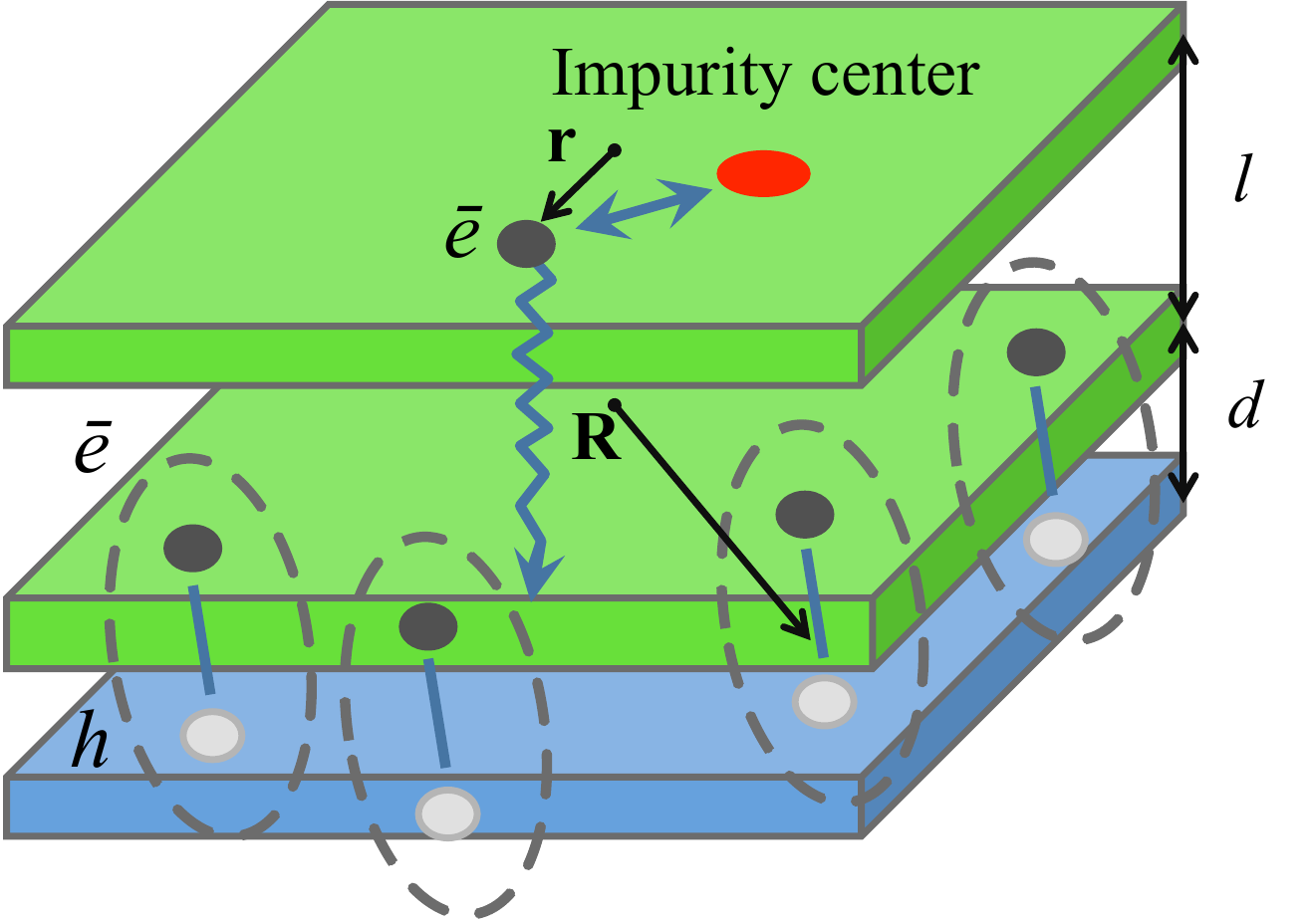}
	\caption{System schematic: spatially separated two-dimensional electron gas (2DEG) with embedded impurity center and a dipolar exciton gas residing in two parallel layers. Charged particles are coupled via the Coulomn interaction.}
			\label{Fig1}
\end{figure}

We consider a hybrid nanostructure consisting of a 2D electron layer separated by a distance $l$ from a double quantum well, containing the dipolar exciton gas, see Fig.~\ref{Fig1}.
The electron-exciton interaction in the 2DEG can be described by the following term in the Hamiltonian:
\begin{gather}\label{eq.1}
V=\int d\textbf{r}\int d\textbf{R}\Psi^\dag(\textbf{r})\Psi(\textbf{r})g(\textbf{r}-\textbf{R})\Phi^\dag(\textbf{R})\Phi(\textbf{R}),
\end{gather}
where $\Psi(\textbf{r})$ and $\Phi(\textbf{R})$ are the quantum field operators of electron and excitons, correspondingly, $g(\textbf{r}-\textbf{R})$ is the Coulomb interaction between an electron and an exciton, $\textbf{r}$ is electron coordinate within the quantum well plane and $\textbf{R}$ is the center of mass exciton coordinate. From now on, we will disregard the internal structure of the excitons and concentrate solely on the density of excitations which represent the collective modes of the exciton gas.

\begin{figure}[!t]
	\includegraphics[width=0.6\linewidth]{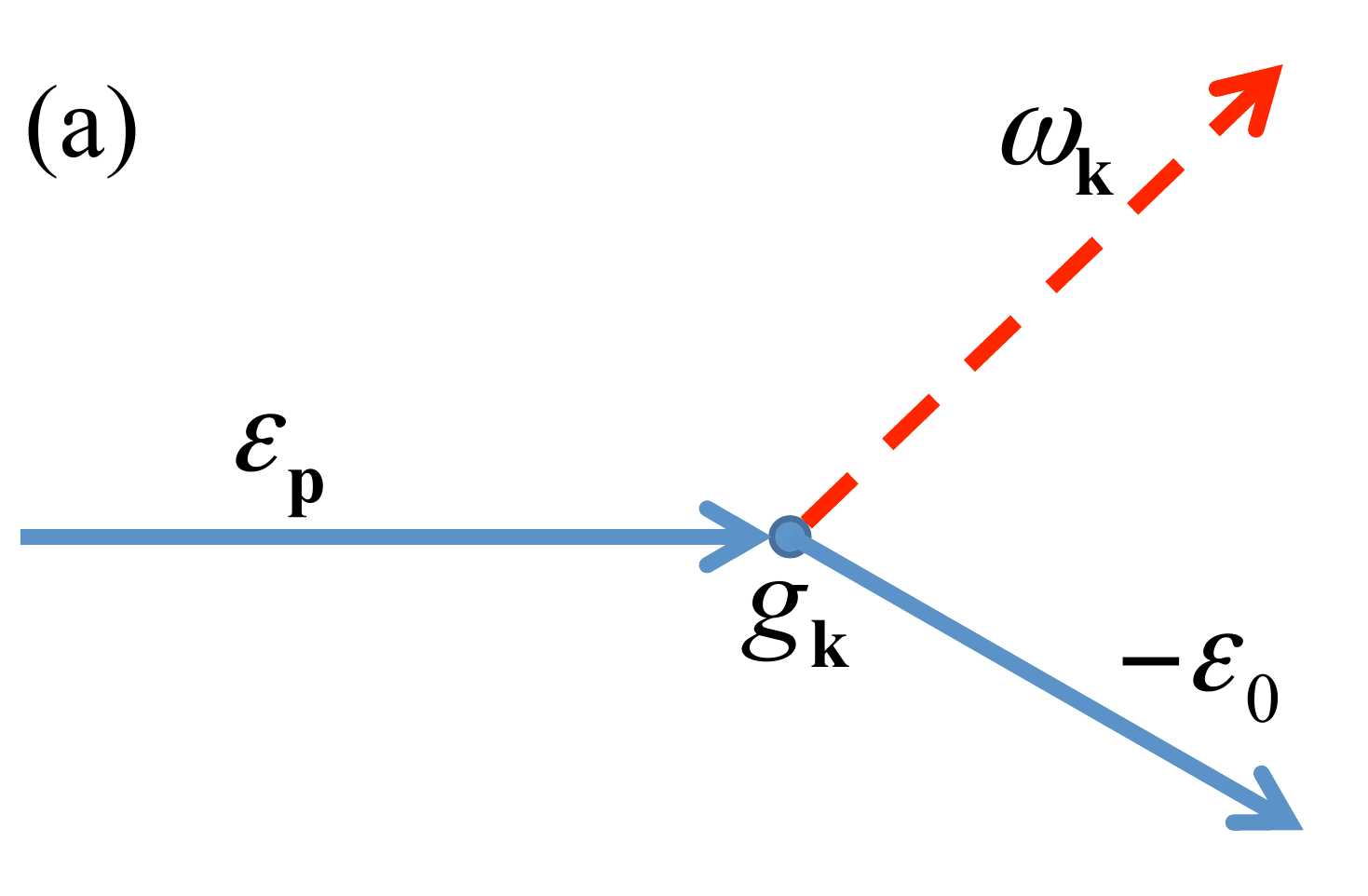}
	\includegraphics[width=0.6\linewidth]{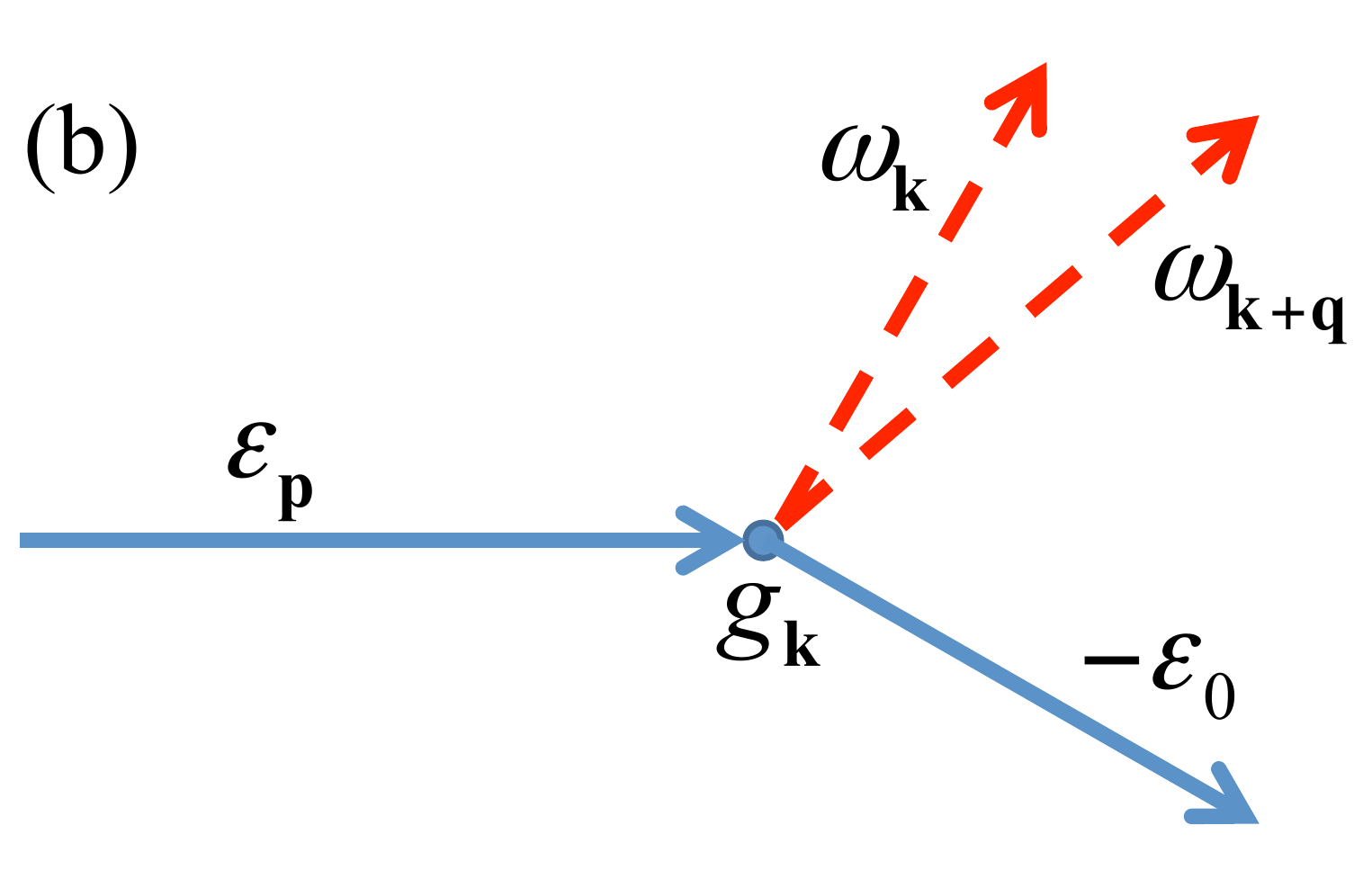}
	\caption{Schematic of the electron capture processes, mediated by the emission of a single (a) and two (b) Bogoliubov quanta (red dashed arrows).}
			\label{Fig2}
\end{figure}

Assuming the exciton gas being in BEC regime, we will use the model of weakly non-ideal Bose gas for their description. The exciton field we present as $\Phi(\textbf{R})=\sqrt{n_c}+\varphi(\textbf{R})$, thus separating the condensed and non-condensed fractions. Here $n_c$ is the exciton condensate density. Then from Eq.~\eqref{eq.1} we yield three contributions:
\begin{gather}\label{eq.2}
V_1=n_c\int d\textbf{r}\Psi^\dag(\textbf{r})\Psi(\textbf{r})\int d\textbf{R}g(\textbf{r}-\textbf{R}),\\\nonumber
V_2=\sqrt{n_c}\int d\textbf{r}\Psi^\dag(\textbf{r})\Psi(\textbf{r})\int d\textbf{R}g(\textbf{r}-\textbf{R})[\varphi^\dag(\textbf{R})+\varphi(\textbf{R})],\\\nonumber
V_3=\int d\textbf{r}\Psi^\dag(\textbf{r})\Psi(\textbf{r})\int d\textbf{R}g(\textbf{r}-\textbf{R})\varphi^\dag(\textbf{R})\varphi(\textbf{R}).
\end{gather}
The operator $V_1$ does not contribute to the electron transition rate due to the energy non-conservation, and therefore it will be further disregarded. Then we take the Fourier transform of the other two operators in~\eqref{eq.2} using the formulae:
\begin{gather}\label{eq.3}
\varphi^\dag(\textbf{R})+\varphi(\textbf{R})=\sum_{\textbf{p}}e^{i\textbf{pR}}\left[(u_\textbf{p}+v_{-\textbf{p}})b_\textbf{p}\right.\\
\nonumber
\left.~~~~~~~~~~~~~~~~~~~~~~~~~~~~~~~~~+(v_\textbf{p}+u_{-\textbf{p}})b^\dag_{-\textbf{p}}\right],\\
\nonumber
\varphi^\dag(\textbf{R})\varphi(\textbf{R})=\sum_{\textbf{p},\textbf{p}'}e^{i(\textbf{p}-\textbf{p}')\textbf{R}}
(u_{\textbf{p}'}b^\dag_{\textbf{p}'}+v_{\textbf{p}'}b_{-\textbf{p}'})\\
\nonumber
~~~~~~~~~~~~~~~~~~~~~~~~~~~~~~~~\times(u_\textbf{p}b_\textbf{p}+v_\textbf{p}b^\dag_{-\textbf{p}}),
\end{gather}
where $b^\dag_{\textbf{p}},b_{\textbf{p}}$ are the creation and annihilation operators of the bogolons, and the coefficients read:
\begin{gather}\label{eq.3.1}
u^2_{\textbf{p}}=1+v^2_{\textbf{p}}=\frac{1}{2}\left(1+\left[1+\frac{(Ms^2)^2}{\omega^2_{\textbf{p}}}\right]^{1/2}\right),\\\nonumber
u_{\textbf{p}}v_{\textbf{p}}=-\frac{Ms^2}{2\omega_{\textbf{p}}}.
\end{gather}
Here $M$ is the exciton mass, $s=\sqrt{\kappa n_c/M}$ is the sound velocity of bogolons, $\kappa=4\pi e^2d/\epsilon$ is exciton--exciton interaction strength, where $d$ is the distance between the layers containing electrons and holes; $\omega_k=sk(1+k^2\xi^2)^{1/2}$ is their spectrum, $\xi=1/(2Ms)$ is the healing length. At small (close to zero) temperature, thermal excitations in the exciton subsystem are suppressed, therefore the processes of electron capture can only be accompanied by the emission of bogolons.
As a result, in Eq.~(\ref{eq.3}) we should concentrate on terms containing ($b^\dag$) and ($b^\dag b^\dag$) only.
Let us consider electron transition from an initial state, $|0_{imp},1_{\textbf{p}}\rangle$, with energy $\varepsilon=\textbf{p}^2/2m$ (zero energy level is taken at the bottom of the lowest electronic subband in the quantum well) to the final bound state, $|1_{imp},0_{\textbf{p}}\rangle$, with energy $-\epsilon_0<0$.
It means that in the electron field operators we keep the terms containing $c^\dag_0$ and $c_\textbf{p}$ only. Then the operators describing single-- and two--bogolon emission processes in momentum representation read:
\begin{gather}\label{eq.4}
V_2=\sqrt{n_c}\sum_{\textbf{k},\textbf{p}}g_\textbf{k}\psi^*_0(\textbf{p}-\textbf{k})(v_{-\textbf{k}}+u_\textbf{k})c^\dag_0c_\textbf{p}b^\dag_\textbf{k},
\end{gather}
\begin{gather}\label{eq.5}
V_3=\sum_{\textbf{k},\textbf{p}}g_\textbf{k}\psi^*_0(\textbf{p}-\textbf{k})c^\dag_0c_\textbf{p}\sum_{\textbf{q}}
u_{\textbf{q}+\textbf{k}}v_{\textbf{q}}b^\dag_{\textbf{q}+\textbf{k}} b^\dag_{-\textbf{q}}.
\end{gather}
In~(\ref{eq.4}) and~(\ref{eq.5}), $g_{\textbf{k}}=2\pi e^2de^{-kl}/\varepsilon$ and $\psi^*_0(\textbf{p})=\int d\textbf{r}e^{-i\textbf{pr}}\psi^*_0(\textbf{r})$ are the Fourier images of the electron--exciton interaction and the wave function of the electron residing at the impurity center, respectively. The schematic of these processes~(\ref{eq.4}) and~(\ref{eq.5}) is presented in Fig.~\ref{Fig2}.
Let us now find the probabilities of the corresponding capture events.


\section{Single--bogolon emission}
The probability of electron capture by the impurity accompanied by the emission of a single bogolon reads:
\begin{eqnarray}\label{eq.6}
\textrm{w}&=&2\pi n_c\sum_{\textbf{k},\textbf{p}}g^2_\textbf{k}|\psi^*_0(\textbf{p}-\textbf{k})|^2
|v_{-\textbf{k}}+u_\textbf{k}|^2
\\
\nonumber
&&~~~\times\delta(\omega_\textbf{k}-\epsilon_0-\textbf{p}^2/2m).
\end{eqnarray}
Here $\omega_\textbf{k}$ is the bogolon dispersion, $m$ is electron effective mass.
It is convenient to make a replacement: $\textbf{p}-\textbf{k}\rightarrow \textbf{p}'$, thus the angle between the two vectors enters the delta-function. Then the integration over the angle can be taken using
\begin{gather}\label{eq.7}
\int_0^{2\pi} d\varphi\delta(a+b\cos\varphi)=2\frac{\theta[|b|-|a|]}{\sqrt{b^2-a^2}}.
\end{gather}
As a result we obtain:
\begin{eqnarray}\label{eq.8}
\textrm{w}&=&\frac{2n_c}{(2\pi)^3}\int kdkg_k^2|v_{-k}+u_k|^2
\\
\nonumber
&&\times\int pdp \frac{|\psi_0^*(p)|^2\theta\left[\frac{pk}{m}-\left|\omega_k-\epsilon_0-\frac{p^2+k^2}{2m}\right|\right]}{\sqrt{\left(\frac{pk}{m}\right)^2-\left(\omega_k-\epsilon_0-\frac{p^2+k^2}{2m}\right)^2}}.
\end{eqnarray}
Due to the presence of $\theta$-function, the limits of integration here should be chosen thus the integrant is positive. In general, Eq.~\eqref{eq.8} requires numerical integration.
However, we can analytically consider the most interesting case corresponding to the slow-electron motion, when $\frac{p^2}{2m}\ll\epsilon_0$. Then we can disregard the kinetic energy of the electron, $\frac{p^2}{2m}$, in the denominator of~\eqref{eq.8}. Besides, let us assume that the electron is captured at the ground state of the Coulomb center, for which we know that $|\psi_0^*(p)|^2=\frac{8\pi a^2}{(1+p^2a^2)^3}$, where $a=\varepsilon\hbar^2/2m e^2$ is a Bohr radius. Integrating over $p$ we find the probability:
\begin{gather}\label{eq.10}
\textrm{w}=\frac{3n_cma}{8\pi}\int_0^\infty \frac{g_k^2|v_{-k}+u_k|^2dk}{\left[1+\frac{m^2a^2}{k^2}\left(\omega_k-\epsilon_0-\frac{k^2}{2m}\right)^2\right]^{5/2}}.
\end{gather}
In the most interesting long--wavelength limit, $k\xi\ll1$, the Bogoliubov quasiparticle dispersion is linear: $\omega_k=sk$. Then Eq.~\eqref{eq.10} can be simplified taking into account that $|v_{-k}+u_k|^2\approx k\xi$ and $\omega_k\gg\frac{k^2}{2m}$. In dimensionless form this equation reads
\begin{gather}\label{eq.11}
\textrm{w}=\frac{3\pi}{8}\left(\frac{d}{a}\right)^2\frac{\epsilon_0^2\xi n_ca}{ms^2}I\left(\frac{e^2l}{\hbar sa};\frac{ma}{2M\hbar\xi}\right),\\\nonumber
I(\alpha;\beta)=\int\limits_{0}^{\infty}\frac{e^{-\alpha x}xdx}{\left[1+\beta^2\left(1-1/x\right)^2\right]^{5/2}}.
\end{gather}
Here we restore the Plank's constant for completeness. This equation is one of the key results of our manuscript.


\section{Two--bogolon emission} The probability of electron capture by the impurity accompanied by the emission of a pair of bogolons reads:
\begin{eqnarray}\label{eq.12}
w&=&2\pi \sum_{\textbf{k},\textbf{p}, \textbf{q}}g^2_\textbf{k}|\psi^*_0(\textbf{p}-\textbf{k})|^2|u_{\textbf{q}+\textbf{k}}v_{\textbf{q}}|^2
\\
\nonumber
&&~~~\times\delta(\omega_{\textbf{q}+\textbf{k}}+\omega_{\textbf{q}}-\epsilon_0-\textbf{p}^2/2m)
\\
\nonumber
&=&2\pi \sum_{\textbf{k},\textbf{p}}g^2_\textbf{k}|\psi^*_0(\textbf{p})|^2\int\limits_{-\infty}^{\infty} d\xi F(\textbf{k},\xi)
\\
\nonumber
&&~~~\times\delta(\xi-\epsilon_0-(\textbf{p}+\textbf{k})^2/2m),
\end{eqnarray}
where we have introduced an auxiliary function:
\begin{gather}\label{eq.13}
F(\textbf{k},\xi)=\sum_{\textbf{q}}|u_{\textbf{q}+\textbf{k}}v_{\textbf{q}}|^2\delta(\xi-\omega_{\textbf{q}+\textbf{k}}-\omega_{\textbf{q}}).
\end{gather}
Integrating in~(\ref{eq.12}) over the angle between $\textbf{p}$ and $\textbf{k}$, we find:
\begin{eqnarray}\label{eq.14}
w&=&\frac{1}{\pi}\sum_{\textbf{k}}g^2_\textbf{k}\int\limits_{-\infty}^{\infty} d\xi F(\textbf{k},\xi)\\\nonumber
&&\times
\int pdp|\psi^*_0(p)|^2\frac{\theta\left[\frac{pk}{m}-\left|\xi-\epsilon_0-\frac{p^2+k^2}{2m}\right|\right]}{\sqrt{\left(\frac{pk}{m}\right)^2-\left(\xi-\epsilon_0-\frac{p^2+k^2}{2m}\right)^2}}.
\end{eqnarray}
Further we assume linear dispersion of the bogolons, $\omega_{\textbf{q}}=sq$, when the coefficients in~\eqref{eq.13} read
$$u_{\textbf{q}+\textbf{k}}\approx\sqrt{\frac{ms}{2|\textbf{q}+\textbf{k}|}},\,\,\,v_\textbf{q}\approx-\sqrt{\frac{ms}{2q}}.$$
According to~(\ref{eq.13}), $\xi\geq sk\gg k^2/2m$, and we again consider a slow electron, $p^2/2m\ll\epsilon_0$. Then, in Eq.~(\ref{eq.14}) one can disregard $\frac{p^2+k^2}{2m}$.
\begin{figure}[!t]
	\includegraphics[width=0.95\linewidth]{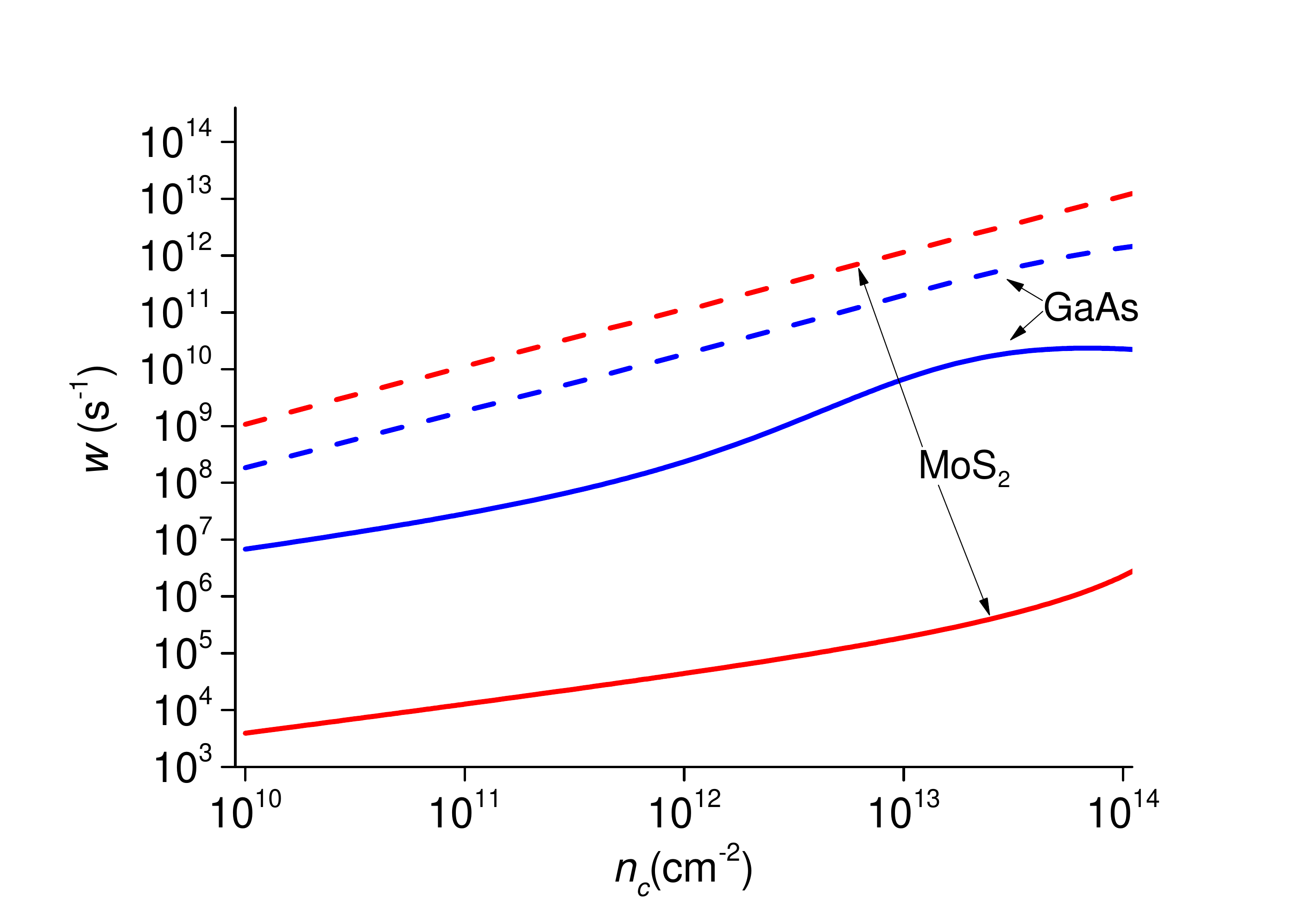}
	\caption{Probability of electron capture as a function of number of particles in BEC, accompanied by the emission of a single bogolon (solid curves) and a pair of bogolons (dashed curves) for GaAs (blue curves) and MoS$_2$ (red curves).}
			\label{Fig3}
\end{figure}
After some derivations, the capture probability takes the dimensionless form:
\begin{eqnarray}
\label{eq.15}
w&=&\frac{3m}{M}\left(\frac{d}{16a}\right)^2\frac{a}{\xi}\frac{\epsilon_0}{\hbar^2}J\left(\frac{e^2l}{\hbar sa};\frac{ma}{2M\hbar\xi}\right),\\
\nonumber
&&J(\alpha;\beta)=\int\limits_{0}^{\infty} \int\limits_{0}^{\infty} \frac{e^{-\alpha x}dxdt}{\left[1+\beta^2\left(\cosh t-1/x\right)^2\right]^{5/2}}.
\end{eqnarray}
This is the second key result of the manuscript.
\begin{figure}[!b]
	\includegraphics[width=1.0\linewidth]{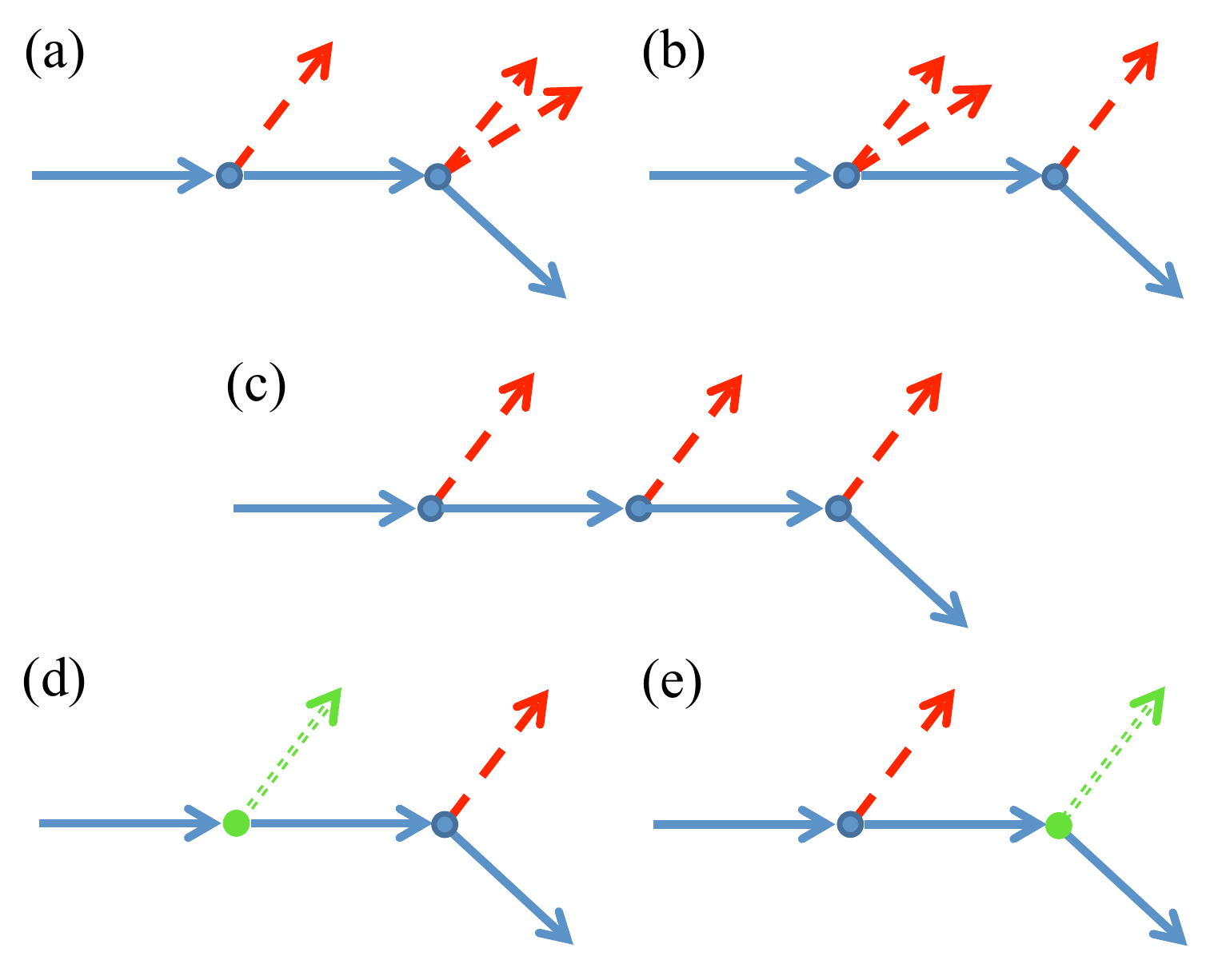}
	\caption{(a-c) Schematic examples (not all the possible diagrams are given here) of the electron capture processes, mediated by the emission of three bogolons, depicted as red dashed arrows, as in Fig.~\ref{Fig2}. (d-e) Schematic examples of electron capture process mediated by the emission of a single bogolon and a single acoustic phonon (not all are given here). Green dotted double lines correspond to phonons and green circles correspond to an electron-phonon interaction bare vertices.
	}
			\label{Fig4}
\end{figure}
%
%
%


\section{Results and discussion}
Figure~\ref{Fig3} shows the difference between the probabilities described by Eq.~(\ref{eq.11}) and Eq.~(\ref{eq.15}). Typical systems where one can observe exciton BEC are based on InAlGaAs and MoS$_2$~\cite{Berman} compounds. We utilize the typical parameters for (i) GaAs nanostructure: $\varepsilon=12.5$, $m=0.067\,m_0$, $M=0.517\,m_0$ ($m_0$ is a free electron mass), $d=10$ nm, $l=50$ nm; and for (ii) MoS$_2$: dielectric constant of h-BN $\varepsilon=4.89$, electron mass $m=0.47\,m_0$, effective mass of $A$-type exciton is $M=0.499\,m_0$, $d=3.5$ nm (about ten monolayers of h-BN) and $l=17.5$ nm~\cite{Kylanpaa}. The bound energy level $\epsilon_0=e^2/\varepsilon a$ within the 2D Hydrogen atom model.

We see from Fig.~\ref{Fig3} that for both materials, the two-bogolon processes are predominant and they cause the capture time to be orders of magnitude less, in comparison with the single bogolon emission events. If for GaAs the difference is of one order of magnitude, for MoS$_2$ the difference is much larger, reaching five orders.
Surprisingly, electron capture events accompanied by emission of a pair of bogolons should be treated within the same order of perturbation theory as the single bogolon emission. If we look back at regular lattice phonons, the probability of emission of a single phonon is proportional to $\alpha_{\textbf{k}}^2$, where $\alpha_{\textbf{k}}$ is the interaction strength. Further, the probability of two--phonon emission contains the factor $\alpha_{\textbf{k}}^4$, indicating the increase in the order of the perturbation theory. Contrast to this, the process of emission of two bogolons in our case has the same order, as a consequence of Coulomb nature of electron-exciton interaction, in contrast with electron-phonon interaction.

%
Another important issue is whether it is correct to disregard the three-- and higher--order bogolon emission processes and if they can be equally important. These processes should be described by the higher-order perturbation theory, and therefore they are much less probable. Several possible diagrams describing the emission of three Bogoliubov quanta are presented in Fig.~\ref{Fig4}a-c. Obviously, these diagrams contain the combination of single-- and two--bogolon emission events, see Fig.~\ref{Fig4}a,b. However, single bogolon emission has much smaller probability amplitudes than the two--bogolon emission, as it was shown above. It results in the decrease of their overall impact, as compared to the diagrams given in Fig.~\ref{Fig2}b. The diagram in Fig.~\ref{Fig4}c gives even smaller contribution being the third--order over the single bogolon emission process.

We would also like to address a `hybrid' case when the capture of the electron is facilitated by simultaneous emission of a bogolon and an acoustic phonon of the crystal lattice. Some of the corresponding diagrams are presented in Fig.~\ref{Fig4}d,e. They contain additional bare electron-phonon vertices (green circles in Fig.~\ref{Fig4}d,e). According to the Migdal's theorem~\cite{Migdal1, Migdal2}, each electron-phonon vertex introduces an additional small factor \textbf{$\sqrt{m/M_a}\ll1$}, where $m$ is the electron mass and $M_a$ is the mass of an atom of the crystal lattice. Thus, the presence of the phonon emission processes increases the order of the perturbation theory of the diagrams and results in the decrease of their impact on the electron capture probability by the small factor $m/M_a$, in comparison with the processes considered in Fig.~\ref{Fig2}. Therefore in the presence of bogolon--mediated electron capture, phonon--assisted processes play a minor role and can be safely disregarded.


\section{Conclusions}
We investigated electron capture by an attractive Coulomb impurity center embedded in a hybrid Bose-Fermi system consisting of spatially separated two-dimensional electron gas and a dipolar exciton BEC gas coupled by the Coulomb interaction. We calculated the probability of electron capture accompanied by the emission of a single bogolon and a pair of bogolons in a single capture event and showed that the latter processes give more important contribution, in contrast with regular acoustic phonon--mediated scattering. As a platform, we studied hybrid systems based on GaAs alloys and MoS$_2$.
We conclude that electron capture by charged impurities in hybrid systems can be strongly enhanced due to the appearance of new type of inelastic scattering processes.

\section*{Acknowledgements}
We thank A. Chaplik for fruitful discussions. We acknowledge the support by the Russian Science Foundation (Project No. 17-12-01039) and the Institute for Basic Science in Korea (Project No.~IBS-R024-D1).

\end{document}